\documentclass{PoS}

\title{30 GHz monitoring of broad absorption line (BAL) quasars}

\ShortTitle{30 GHz monitoring of broad absorption line (BAL) quasars}
\author{\speaker{Maciej Ceglowski}%
         \\
        Torun Centre for Astronomy, Gagarina 11, 87-100 Torun, Poland\\
        E-mail: \email{ceglowski@astri.uni.torun.pl}}

\author{Magdalena Kunert-Bajraszewska\\
        Torun Centre for Astronomy, Gagarina 11, 87-100 Torun, Poland\\
        E-mail: \email{magda@astro.uni.torun.pl}}
        \author{Bogna Pazderska\\
        Torun Centre for Astronomy, Gagarina 11, 87-100 Torun, Poland\\
        E-mail: \email{bogna@astro.uni.torun.pl}}
\author{Marcin Gawronski\\
        Torun Centre for Astronomy, Gagarina 11, 87-100 Torun, Poland\\
        E-mail: \email{motylek@astro.uni.torun.pl}}
\abstract{
Broad absorption line (BAL) quasars have been studied for over thirty years. 
Yet it is still unclear why and when we observe broad absorption lines in
quasars. Is this phenomenon caused by geometry  or is it 
connected with the
evolution process? Variability of the BAL quasars, if present, can give us
information about their orientation, namely it can indicate whether they are
oriented more pole-on. Using the Torun 32-metre dish equipped with the One Centimetre
Receiver Array (OCRA) we have started a monitoring campaign of a sample of
compact radio-loud BAL quasars. This 30 GHz variability monitoring program 
supplements the high-resolution interferometric observations of these objects 
we have
carried out with the EVN and VLBA.   
}
\FullConference{11th European VLBI Network Symposium \& Users Meeting\\
                 9-12 October 2012\\
                 Bordeaux (France)}

\begin{document}

\section{Introduction}

Broad absorption lines (BALs) are seen in a small (15 \%) fraction of both the radio-quiet and 
radio-loud quasar populations \cite{knig}, according to the traditional BALQSO definition 
of \cite{wey}. They are probably caused by outflow of gas with high velocities and are
part of the accretion process.  The presence of BALs is probably a geometrical effect 
\cite{elvis00} but it can be also connected with quasar evolution \cite{gregg00}.
Theoretical models suggest that BALs are seen at high inclination angles, which means that the
outflows from accretion disks are present near the equatorial plane.
However, there are also observational evidences indicating the existence of polar outflows from the
inner regions of a thin disk \cite{zhou06,ghosh07}. This means that
either there is not a simple orientation model which can explain all the
features observed in BALQSOs and/or the evolution scenario should be taken into
account.

The evolution scenario which emerged with the discovery of radio-loud
BALQSOs suggests that every quasar has a BAL phase at the beginning of
its lifetime \cite{becker00}. Recent analysis of the spectral shape,
variability and polarization properties of some of the compact radio-loud
BALQSOs \cite{monte08} indicates that indeed
they are similar to young compact steep spectrum
(CSS) and gigahertz peaked spectrum (GPS) objects.

The VLBI technique is still the best way to learn about morphologies and
orientation of BAL quasars. However, only a very small
fraction of compact BAL quasars have been observed with VLBI so far
\cite{jiang03,liu08,monte08,kunert09,doi09,kunert10,bruni13, hayashi}.
About half of the observed sources have unresolved structures which
prevents us from directly estimating their orientation from morphology, but
the steepness of their radio spectrum can give us an idea about their
orientation.
Additionally, BAL quasars with polar rather than equatorial outflows have   
likely been identified via short timescale variability \cite{zhou06}.    

A few years ago we started  high-resolution EVN and VLBA observations of selected 
samples of compact radio-loud broad absorption line quasars to study their morphology
\cite{kunert10pos}.  
Last year we have also initiated a flux variability monitoring campaign of these samples with 
the One Centimetre Receiver Array (OCRA) prototype system mounted 
on the Torun 32 m radiotelescope.

\section{OCRA receiver}
The OCRA project aims to construct a one-hundred-beam receiver 
system, operating in a frequency band centred on 30 GHz \cite{browne00}. This system is
to be installed on the 32m parabolic antenna at Torun Observatory. 
The antenna is ideal for high-frequency multi-beam work as it has very accurate reflector panels and paraboloid/hyperboloid 
(non-shaped) optics. Having 100 beams would allow all-sky surveys to be carried out much
quicker than is currently possible. Currently there are two ongoing projects,
one designated OCRA-p and the other
OCRA-F. They are being used to develop the technology needed for the full OCRA system.
OCRA-p is a two-beam prototype receiver, based on the design of the 30 GHz Planck Low Frequency Instrument, and 
OCRA-F is an 8-beam receiver using the MMICs technology \cite{pengelly75}. Our monitoring of BAL quasars is
carried out with OCRA-p. The sensitivity and stability of the receiver
allow us to observe sources as weak as 10\,mJy, with improvements still going on.

\section{Sample and observations}
We have cross-identified the FIRST survey with a catalog of optically-selected BAL
quasars from SDSS/DR3 \cite{trump06} with the following selection
criteria: a) unresolved, isolated sources i.e more compact than the FIRST
beam (5.4 arcsec), b) sources surrounded by an empty field within a radius of 1
arcmin, c) sources with a flux density $S_{1.4\,{\rm GHz}}>20\,{\rm
mJy}$. The final sample consist of 26 compact radio-loud BAL quasars.

We have already observed part of the sample with the VLBA at 5 and 8.4\,GHz in polarization mode.
The remaining objects of the sample are currently being observed 
with the EVN at 5\,GHz. Most of the sources that have been observed so far
reveal core-jet structures and show fainter emission from the jet compared to the core. This morphology 
suggests intermediate orientation between polar and equatorial geometries and 
supports the orientation scenario. For some sources the flux densities of the observed compact structures
account  only for up to 25\% of the total flux density
at 5\,GHz. This suggests that these sources comprise low brightness
extended structures and  may
be older and bigger than GPS/CSS objects \cite{kunert10pos}.

The 30\,GHz monitoring of the whole sample is being carried out
using OCRA-p in "on-off" mode. Each source is observed once per month and each flux measurement 
is done twice. We have divided our sample into two groups of objects
based on their 1.4\,GHz flux density. The $S_1$ group consists of the 16 brighter sources
with flux densities $S_{1.4\,{\rm GHz}}>150\,{\rm mJy}$. The second group
$S_2$
consists of 10 much weaker objects with flux densities $S_{1.4\,{\rm
GHz}}>20\,{\rm mJy}$. 
14 out of the     
16 sources from the $S_1$ group have been detected at 30 GHz so far,and their flux density values range from  9.5 mJy   
to 270\,mJy. Their spectral indices between 1.4 GHz and  5 GHz range from 
--0.33 to 0.9 and get steeper between 5\,GHz and 30\,GHz. Sources
 belonging to the $S_2$ group have flux density that is more than 3 times weaker   
and only 3 of them have been detected at 30\,GHz so far.
Weather conditions play an important role as far as data quality is concerned. 
However, with the monitoring going on, this limitation will be overcome as more data are acquired.

By carrying out this project we intend to address several important issues, 
one of which is to find out the BAL quasar orientation. 
Short timescale radio flux variations of BAL\,QSOs  
may help to confirm the previous findings that viewing angles are within $35^{\rm o}$ of the jet axis
\cite{ghosh07}. From a preliminary analysis of our high-resolution VLBI
observations, supplemented with those found in the litterature, 
we suspect that flux and polarisation variability is present in some of
them. 
Our second goal is to obtain the spectral energy distribution (SED) of these BALQSOs. 
This will allow us to study their spectral shape, namely its steepness, and
check the potential presence of low brightness extended structures.

\bigskip
\noindent
{\small
{\bf Acknowledgements}\\
\noindent
This work was supported by the Polish Ministry of Science and
Higher Education under grant UMO-2011/01/D/ST9/00378.
We gratefully acknowledge the financial support of the Royal Society Paul
Instrument Fund which allowed us to build the 30 GHz OCRA receiver. We are
also grateful to the Polish Ministry of Science and Higher Education
which provided support for operating the
receiver on the Torun 32-m telescope.}

\end{document}